\documentclass[a4paper]{JHEP3}
\usepackage{amsmath}
\usepackage{amsfonts}
\usepackage{amssymb}
\usepackage{epsfig}
\usepackage{graphicx}
\usepackage{amssymb}
\usepackage{latexsym}

\newcommand{\be}{\begin{equation}}
\newcommand{\ee}{\end{equation}}
\newcommand{\bea}{\begin{eqnarray}}
\newcommand{\eea}{\end{eqnarray}}

\newcommand{\del}{\partial}
\newcommand{\td}{\textup{d}}  
\def\eqref#1{(\ref{#1})}
\def\er#1{eqn.\eqref{#1}}
\def\nn{\nonumber}

\title{Spectrum to all orders of Polchinski-Strominger {Effective} String Theories of  the Drummond Type}
\author{N.D. Hari Dass
\\ CHEP , Indian Institute of Science, Bangalore 560012, India \footnote{DAE Raja Ramanna Fellow} \\
Poornaprajna Institute of Scientific Research, Bangalore 560080, India\\
 Email: \email{dass@cts.iisc.ernet.in}}
\author{Yashas Bharadwaj
\\ CHEP , Indian Institute of Science, Bangalore 560012, India \\
 Email: \email{bharadwajy@cts.iisc.ernet.in}}

\abstract{
Drummond had proposed four actions for Polchinski-Strominger effective string theories at order $R^{-6}$, where
$2\pi R$ is the length of the (closed) string . In \cite{covariant} it had been
shown, based on covariance arguments, that only two of them are independent. We analyse the spectral content of 
effective string theories with these two actions. We show that the inclusion of these actions does not yield 
corrections to the spectrum of
Nambu-Goto theory \cite{Arv}. 
}


\begin{document}

\section{Introduction}

Effective string theories are of interest as consistent ways of quantum mechanically describing string-like defects.
Of particular interest are those circumstances where there are only \emph{massless transverse} degrees of freedom.
Two approaches to such effective string theories exist in the literature. One due to L\"uscher and collaborators 
\cite{lwearly,lwrecent}, 
is formulated
entirely in terms of the $D-2$ \emph{transverse} degrees of freedom. It is a case where the \emph{gauge} is fixed completely
without any \emph{residual} invariance left. 
The recent work of Aharony and Karzbrun \cite{aharony} has followed this approach in addressing
the issue of spectrum of effective string theories to higher orders. We, along with Drummond, have on the other hand followed
the approach pioneered by Polchinski and Strominger \cite{ps}. In the latter approach, the theories are invariant under \emph{conformal
transformations} and the physical states are obtained by requiring that the generators of conformal transformations
annihilate them. These too are gauge-fixed theories but with leftover residual invariances characterized by conformal
transformations. It is worth emphasizing that the physical basis of both approaches is that the degrees of freedom are 
transverse.

In their pioneering work Polchinski and Strominger showed how to quantise effective string theories consistently in all
dimensions, and also gave an algorithm to construct actions beyond what they had studied in \cite{ps}. Drummond and, 
Hari Dass and Matlock,
developed further the systematics of such a construction \cite{drumorig,ouruniv}. The upshot of their analysis was 
the absence of any candidate actions at $R^{-3}$ level 
 (where $2\pi R$ is the length of the string) 
beyond what had already been considered in \cite{ps}. Polchinski and Strominger had already 
stated in \cite{ps}, without proof, that corrections to their action was only expected at $R^{-4}$. A more striking result of 
\cite{drumorig,ouruniv} was the \emph{universality} in the spectrum of effective string theories to order $R^{-3}$; by this we mean
the equality of the spectrum of all effective string theories to this order with that of the free bosonic string theory \cite{Arv}. Even more
surprisingly, Drummond had shown that there were additional candidate actions
only at the $R^{-6}$ order.
He explicitly wrote down four action terms. The transformation laws leaving those actions invariant was not addressed by him. Subsequently we \cite{covariant} formulated a \emph{Covariant Calculus} to construct
actions where the transformation laws remained the same as in the free theory. In \cite{covariant} we also showed how
to covariantize what we have called  the Drummond actions and showed that only two of them were \emph{independent}.

In this paper we analyse these independent actions for their influence on the spectrum of effective string theories.
This is an issue of great importance to understand the nature of QCD-strings \cite{bali,kutilat05,lwrecent,ournum}. Our own interest in effective string theories
was kindled by our high accuracy numerical simulations of the static antiquark-quark potentials in three and four dimensions
where we showed evidence that to good accuracy it agreed with the ground state energy of a free bosonic string theory to
order $R^{-3}$\cite{ournum}.
This surprising result was analytically proved in \cite{drumorig,ouruniv} in the Polchinski-Strominger approach. Recently Aharony and Karzbrun have shown this result to be true following the L\"uscher approach. At the same time we \cite{liou-all} 
have shown that a
generalization of the PS-action which is exactly conformally invariant to \emph{all orders} also does not correct 
the Nambu-Goto
result to \emph{all orders}. As a step towards studying similar results for \emph{all} effective string theories, we analyse the
Drummond actions here. We show that they too do not change the spectrum from that of the free bosonic string theory.
\section{ Covariantising the Drummond Actions }
\label{covdrum}
\subsection{General Considerations}
\label{conf1}
We start with the form of manifestly general covariant action terms, more
specifically, terms that transform as scalar densities. A systematic
procedure for construction of such terms to any desired order in $1/R$
is given in \cite{covariant}.
\begin{equation}
I_{\textup{cov}} = \sqrt{g}D_{\alpha_1\beta_1..}X^{\mu_1}D_{\alpha_2\beta_2..}X^{\mu_2}\cdot A^{\alpha_1\beta_1\cdots\alpha_2\beta_2\cdots}B_{\mu_1\mu_2\cdots}
\end{equation}
where $A^{\alpha_1\beta_1\cdots\alpha_2\beta_2\cdots}$ is composed of
suitable factors of Levi-Civita and metric tensors on the
two-dimensional world sheet and $B_{\mu_1\mu_2\cdot}$ made up of
$\eta_{\mu\nu}$ and Levi-Civita tensors in target space. In the spirit of
the PS-construction, the covariant calculus is constructed based on the \emph{induced metric}
on the world-sheet given by $g_{\alpha\beta} = \del_\alpha X\cdot \del_\beta X$.
In the
conformal gauge, $g_{++} = g_{--} =0$,
this construction can be done even more simply by
stringing together a number of covariant derivatives so that there are
equal net numbers of $(+,-)$ indices, and finally use sufficient
inverse powers of $g_{+-}\equiv L = \del_+X\cdot\del_-X$ to make the expression transform as $(1,1)$.
The residual transformations maintaining the conformal gauge result in the exact invariance of these actions
under
\begin{equation}
\label{covtrans}
\delta_\pm~X^\mu = -\epsilon^\pm(\tau^\pm)\del_\pm X^\mu
\end{equation}
In this gauge $g_{+-}=g_{-+}=L$ transforms as a true $(1,1)$-tensor
under these conformal transformations, and, 
$g^{+-}=g^{-+}=L^{-1}$ as a $(-1,-1)$ tensor.  
The non-vanishing components of the
Christoffel connection 
are:
\begin{equation}
\label{Gamma1}
{\Gamma^{(1)}}^+_{++} = \del_+ \ln L;~~~~{\Gamma^{(1)}}^-_{--} = \del_- \ln L
\end{equation}
We give explicit expressions for some covariant derivatives of interest to this paper:
\begin{eqnarray}
\label{covderivs}
D_\pm~X^\mu &=& \del_\pm~X^\mu\nn\\
D_{++}X^\mu &=& \del_{++}X^\mu - \del_+\ln L\del_+ X^\mu\nn\\
D_{--}X^\mu& =& \del_{--}X^\mu - \del_-\ln L\del_-X^\mu\nn\\
D_{+-} X^\mu &=& D_{-+}X^\mu = \del_{+-}X^\mu\nn\\
D_{++-} X^\mu &=& D_{+-+} X^\mu = \del_{++-} X^\mu -\del_+\ln L~\del_{+-} X^\mu\nn\\
D_{-++} X^\mu &=& \del_{-++} X^\mu - \del_-(\del_+\ln L~\del_{+} X^\mu)  
\end{eqnarray}

Drummond \cite{drumorig} found four possibilities for effective Lagrangeans at order
$R^{-6}$. 

\begin{eqnarray}
\label{dterms1}
{\cal L}^D_1 & = & \frac{1}{L^3} \del_+^2X\cdot\del_+^2X~\del_-^2X\cdot\del_-^2 X \\
\label{dterms2}
{\cal L}^D_2 & = & \frac{1}{L^3} \del_+^2X\cdot\del_-^2X~\del_+^2X\cdot\del_-^2 X \\
\label{dterms3}
{\cal L}^D_3 &=& \frac{1}{L^4} \del_-^2 X \cdot \del_+^2 X \del_- X \cdot \del_+^2 X \del_-^2 X \cdot \del_+ X, \\
\label{dterms4}
{\cal L}^D_4 &=& \frac{1}{L^5} (\del_- X \cdot \del_+^2 X)^2 (\del_-^2 X \cdot \del_+ X)^2
.\end{eqnarray}

\subsection{Covariantising the Drummond Terms}
\label{covIdrum}
The modified conformal transformations that leave these invariant was not determined in \cite{drumorig}.
Inspection of the first two terms indicates that  one can expect these to be contained in
the covariant forms
\begin{equation}
\label{cov1ex1}
{\cal M}_1  =  \sqrt{g}D_{\alpha_1\beta_1}X\cdot D_{\alpha_2\beta_2}X~D^{\alpha_1\beta_1}X\cdot D^{\alpha_2\beta_2}X 
\end{equation}
\begin{equation}
\label{cov1ex2}
{\cal M}_2  =  \sqrt{g}D_{\alpha_1\beta_1}X\cdot D^{\alpha_1\beta_1}X~D_{\alpha_2\beta_2}X\cdot D^{\alpha_2\beta_2}X
\end{equation}

Let us start with \er{cov1ex1} and \er{cov1ex2}. It is easy to work out 
these expressions in the conformal gauge.
\begin{eqnarray}
{\cal M}_1 &=& 2 \frac{D_{++}X\cdot D_{++}X~D_{--}X\cdot D_{--}X}{L^3} 
+ 2\frac{(D_{++}X\cdot D_{--}X)^2}{L^3}\nn\\
\end{eqnarray}
\begin{equation}
{\cal M}_2 = \frac{4}{L^3}(D_{++}X\cdot D_{--}X)^2
\end{equation}
We consider the particular combination
\begin{equation}
{\cal M}_1-\frac{{\cal M}_2}{2} = \frac{2}{L^3}(D_{++}X\cdot D_{++}X)(D_{--}X\cdot D_{--}X)
,\end{equation}
and it is easy to show that, modulo terms that are leading-order
constraints $\del_{\pm}X\cdot\del_{\pm}X$  and their derivatives, this is just ${\cal L}_1^D$. To
understand ${\cal M}_2$, we note that on using 
\begin{equation}
\label{prel3}
L^{-1}D_{++}X\cdot D_{--}X = \del_-(L^{-1}\del_+^2X\cdot\del_-X) + \textup{EOM}
\end{equation}
It follows that
\begin{eqnarray}
{\cal M}_2 &=& L^{-1}[L^{-1}\del_+^2X\cdot\del_-^2X-L^{-2}\del_-L~\del_+^2X\cdot\del_-X]^2\nn\\
&=& {\cal L}^D_2-2{\cal L}^D_3 +{\cal L}^D_4
\end{eqnarray}
This way we are able to obtain two independent linear combinations of \er{dterms1}.
It can be shown, through straightforward but tedious algebra, that the covariant
calculus can not produce any other combinations. The obvious approach to covariantising
the rest of \er{dterms1} by replacing ordinary derivatives by covariant derivatives only
produces, apart from these combinations, EOM and derivatives, constraints and their derivatives, and
total derivatives. 
\section{Analysis of Covariant Drummond Actions}
\label{covdrumanalysis}
\subsection{Stress Tensors}
Let us consider the following linear combination of actions
\begin{eqnarray}
\label{covdrums}
S^D &=& ~S_1^D + S_2^D\nn\\
 &=& \frac{\eta_1}{4\pi}~\int \td\tau^+\td\tau^-{\cal M}^D_1 + 
  \frac{\eta_2}{4\pi}~\int \td\tau^+\td\tau^-{\cal M}^D_1  
\end{eqnarray}
where
\begin{eqnarray}
\label{drumlags}
{\cal M}_1^D &=&
{\frac{{\left(D_{++}X \cdot D_{--}X \right)}^{2}}{L^{3}}}\\  
{\cal M}_2^D &=&
{ \frac{\left(D_{++}X \cdot D_{++}X\right) \left( D_{--}X \cdot D_{--}X \right)}{L^{3}}} 
\end{eqnarray}

It is to be understood that \er{covdrums} is always accompanied by the action of the free bosonic string theory (Nambu-Goto
action)
\begin{equation}
\label{nambugoto}
S_0 = \frac{1}{4\pi a^2}\int\td\tau^+\td\tau^- \del_+X\cdot\del_-X
\end{equation}
as well as by, what we have called the \emph{Polyakov-Liouville} action \cite{liou-all}
\begin{equation}
\label{liouville}
S_{(2)} = \frac{\beta}{4\pi}\int \td\tau^+\td\tau^-~\frac{\del_+ L\del_- L}{L^2}
\end{equation}
The first provides the leading order action while the second, as shown by Polchinski-Strominger \cite{ps} and \cite{liou-all}, 
provides quantum
consistency in all dimensions. Let us first consider the variation of $S_1^D$ under \er{covtrans}. After a lot of tedious algebra it follows that

\begin{equation}
\label{drumlag1}
\delta_-{\cal M}_1^D = \partial_{-} \left\lbrace \epsilon^{-} \frac{1}{L^{3}} {\left(D_{++}X \cdot D_{--}X \right)}^{2} \right\rbrace
=\del_-(\epsilon^-~{\cal M}_1^D)
\end{equation}
\er{drumlag1} shows that ${\cal M}_1^D$ transforms like a scalar density. 

To obtain the stress tensor $T_{--}$, the generator of the transformation $\delta_- X$, we follow the N\"oether procedure
wherein we consider variations of actions under \er{covtrans} when $\epsilon^-$ is taken to depend on both $(\tau^+,\tau^-)$
(likewise for $\epsilon^+$). The $T_{--}$ is then defined as
\begin{equation}
\label{stressdef}
\delta S = \frac{1}{2\pi}\int \partial_{+} \epsilon^{-} T_{--}^{\left(1\right)} d\tau^{+} d\tau^{-}.
\end{equation}
After considerable algebra it follows that
\begin{eqnarray}
\delta S_{1}^D &=& \frac{\eta_{1}}{4 \pi} \int \partial_{+} \epsilon^{-} \Bigg\lbrace - 
\frac{2}{L^{3}} D_{+} \left[ \left(D_{++}X \cdot D_{--}X \right)D_{-}X \cdot D_{--}X \right]\nn\\ 
 &+& \frac{2}{L^{4}}D_{+} \left[ \left(D_{++}X \cdot D_{--}X \right)  D_{-}X \cdot D_{-} X \left( D_{+}X \cdot D_{--}X 
\right) \right]  + \frac{2}{L^{3}}D_{-} \left[  \left(D_{++}X \cdot D_{--}X \right)\left( D_{+}X \cdot D_{--}X 
\right)\right] \nn\\ 
 &+& \frac{4}{L^{3}} \left(D_{++}X \cdot D_{--}X \right)D_{+-}X \cdot D_{--}X - \frac{4}{L^{4}} \left(D_{++}X \cdot D_{--}X \right)  D_{+-}X \cdot D_{-} X \left( D_{+}X \cdot D_{--}X \right)\nn\\   
 &+&\frac{2}{L^{4}} D_{-} \left[\left(D_{++}X \cdot D_{--}X \right) D_{-}X \cdot D_{-} X\left( D_{++}X \cdot D_{-}X 
\right)\right] \nn\\ 
&-&\frac{4}{L^{4}} \left(D_{++}X \cdot D_{--}X \right) D_{--}X \cdot D_{-} X \left( D_{++}X \cdot D_{-}X \right)
-\frac{3}{L^{4}} \left( D_{-}X \cdot D_{-}X \right){\left(D_{++}X \cdot D_{--}X \right)}^{2}\Bigg\rbrace  
\end{eqnarray}

Therefore, 
\begin{eqnarray}
\label{stressd1}
T_{--}^{D1} & =& \frac{\eta_{1}}{2 } \Bigg\lbrace - \frac{2}{L^{3}} D_{+} \left[ \left(D_{++}X \cdot D_{--}X \right)D_{-}X 
\cdot D_{--}X \right]\nn\\  
 &+& \frac{2}{L^{4}}D_{+} \left[ \left(D_{++}X \cdot D_{--}X \right)  D_{-}X \cdot D_{-} X \left( D_{+}X \cdot D_{--}X 
\right) \right]  + \frac{2}{L^{3}}D_{-} \left[  \left(D_{++}X \cdot D_{--}X \right)\left( D_{+}X 
\cdot D_{--}X \right)\right] \nn\\  
 &+& \frac{4}{L^{3}} \left(D_{++}X \cdot D_{--}X \right)D_{+-}X \cdot D_{--}X - \frac{4}{L^{4}} \left(D_{++}X 
\cdot D_{--}X \right)  D_{+-}X \cdot D_{-} X \left( D_{+}X \cdot D_{--}X \right) \nn\\   
 &+&\frac{2}{L^{4}} D_{-} \left[\left(D_{++}X \cdot D_{--}X \right) D_{-}X \cdot D_{-} X\left( D_{++}X 
\cdot D_{-}X \right)\right] \nn\\  
&-&\frac{4}{L^{4}} \left(D_{++}X \cdot D_{--}X \right) D_{--}X \cdot D_{-} X \left( D_{++}X \cdot D_{-}X \right)
-\frac{3}{L^{4}} \left( D_{-}X \cdot D_{-}X \right){\left(D_{++}X \cdot D_{--}X \right)}^{2}\Bigg\rbrace    
\end{eqnarray}

It likewise follows after some work that
\begin{equation}
\delta{\cal M}_2^D = \del_-(\epsilon^-~{\cal M}_2^D)
\end{equation}
under \er{covtrans} and the N\"oether variation is
\begin{eqnarray}
\delta S_{2}^D &=& \frac{\eta_{2}} {4 \pi} \int\partial_{+} \epsilon^{-} \Bigg\lbrace  -\frac{2}{L^{3}}D_{+} \left[ \left(D_{--}X \cdot D_{--}X \right)D_{-}X \cdot D_{++}X \right]\nn\\ 
 &+&\frac{2}{L^{4}} D_{+}\left[ \left(D_{--}X \cdot D_{--}X \right) \left( D_{-}X \cdot D_{-} X \right)D_{+}X 
\cdot D_{++}X \right] +  \frac{2}{L^{3}} D_{-} \left[\left(D_{--}X \cdot D_{--}X \right)\left( D_{+}X \cdot D_{++}X 
\right) \right]\nn\\ 
 &+&\frac{4}{L^{3}} \left(D_{--}X \cdot D_{--}X \right) D_{+-}X \cdot D_{++}X -\frac{4}{L^{4}} \left(D_{--}X 
\cdot D_{--}X \right) D_{+-}X \cdot D_{-} X \left( D_{+}X \cdot D_{++}X \right)\nn\\  
 &+&\frac{2}{L^{4}} D_{-} \left[\left(D_{++}X \cdot D_{++}X \right) D_{-}X \cdot D_{-} X\left( D_{--}X \cdot D_{-}X 
\right)\right]\nn\\ 
&-&\frac{4}{L^{4}} \left(D_{++}X \cdot D_{++}X \right) D_{--}X \cdot D_{-} X 
 \left( D_{--}X \cdot D_{-}X \right)\nn\\
&-& \frac{3}{L^{4}}\left( D_{-}X \cdot D_{-}X \right)\left(D_{++}X 
\cdot D_{++}X \right) \left(D_{--}X \cdot D_{--}X \right)\Bigg\rbrace 
 \end{eqnarray}
 Leading to 
 \begin{eqnarray}
\label{stressd2}
 T_{--}^{D2} &=& \frac{\eta_{2}} {2 }\Bigg\lbrace  -\frac{2}{L^{3}}D_{+} \left[ \left(D_{--}X \cdot D_{--}X \right)D_{-}X 
\cdot D_{++}X \right]
+  \frac{2}{L^{3}} D_{-} \left[\left(D_{--}X \cdot D_{--}X \right)\left( D_{+}X 
\cdot D_{++}X \right) \right] \nn\\ 
& +&\frac{2}{L^{4}} D_{+}\left[ \left(D_{--}X \cdot D_{--}X \right) \left( D_{-}X \cdot D_{-} X \right)D_{+}X 
\cdot D_{++}X \right] \nn\\
&+&\frac{2}{L^{4}} D_{-} \left[\left(D_{++}X \cdot D_{++}X \right) D_{-}X \cdot D_{-} X\left( D_{--}X 
\cdot D_{-}X \right)\right] \nn\\ 
& -&\frac{4}{L^{4}} \left(D_{--}X \cdot D_{--}X \right) D_{+-}X \cdot D_{-} X \left( D_{+}X 
\cdot D_{++}X \right) \nn\\
& -&
\frac{4}{L^{4}} \left(D_{++}X \cdot D_{++}X \right) D_{--}X \cdot D_{-} X 
 \left( D_{--}X \cdot D_{-}X \right)\nn\\
& -& \frac{3}{L^{4}}\left( D_{-}X \cdot D_{-}X \right)\left(D_{++}X 
\cdot D_{++}X \right) \left(D_{--}X \cdot D_{--}X \right)
+\frac{4}{L^{3}} \left(D_{--}X \cdot D_{--}X \right) D_{+-}X \cdot D_{++}X \Bigg\rbrace\nn\\
 \end{eqnarray}

Another very important equation obeyed by the covariant stress tensors \er{stressd1} and \er{stressd2} is
\begin{equation}
\label{off-eom}
D_+~T_{--} = \del_+~T_{--} = 2\pi E\cdot D_-X
\end{equation}
where $E^\mu = \frac{\delta S}{\delta X^\mu}$ are the equations of motion. The explicit form of $E^\mu$ is not needed.

\subsection{Analysis of Spectrum}

Following \cite{ps,liou-all} we first need to calculate the \emph{on-shell} stress tensor which, by \er{off-eom}, is
\emph{holomorphic}. To investigate the holomorphic content of the stress tensors \er{stressd1} and \er{stressd2} 
\emph{on-shell} we introduce, as in \cite{liou-all}, the decomposition
\begin{equation}
\label{holosplit}
X^\mu = X^\mu_{cl}+F^\mu(\tau^+)+G^\mu(\tau^-)+H^\mu(\tau^+,\tau^-)
\end{equation}
where $H^\mu$ is \emph{purely non-holomorphic} in the sense that by construction it is free of  purely holomorphic or
purely antiholomorphic parts. In \er{holosplit}, $X^\mu_{cl}=R(e_+^\mu~\tau^++e_-^\mu~\tau^-)$ is a classical solution
of the leading order EOM of $S_0$. Because of the linearity of the boundary conditions it follows that $F,G$ are polynomial-free.
  This construction allows all tensors to be likewise split into holomorphic, antiholomorphic
and non-holomorphic pieces. Therefore, splitting the stress tensor $T_{--}$ into $T_{--}=T_{--}^h+T_{--}^{nh}$ where
$T_{--}^h$ is purely holomorphic and $T_{--}^{nh}$ all the rest, it follows from \er{off-eom} that on-shell
$T_{--}^{nh}=0$. 
 
Upon using \er{covderivs} and \er{holosplit} we find
\begin{equation}
D_{++}X^{\mu} = \left(F_{++}^{\mu}+H_{++}^{\mu}\right)-\partial_{+}\left(\ln L\right)\left(Re_{+}^{\mu}
+F_{+}^{\mu}+H_{+}^{\mu}\right) 
\end{equation}
But
\begin{equation}
L = -\frac{R^2}{2}+e_+\cdot(G_-+H_-)+e_-\cdot(F_++H_+)+(G_-+H_-)\cdot(F_++H_+)
\end{equation}
Making $\del_+L$ nonholomorphic. This means that $D_{++}X^{\mu}$ is non-holomorphic. Hence, it's derivatives will also be non-holomorphic.
It then follows that every term in \er{stressd1} as well as in \er{stressd2} is non-holomorphic. Consequently the covariantised Drummond actions do not contribute \emph{anything} to the on-shell stress tensor of the effective theory. Explicitly
stated, the spectrum of effective string theories containing the Drummond terms are again the same as the spectrum of the Nambu-Goto theory
\section{Discussion and Conclusions}
In this paper we have firstly complemented Drummond's construction of $R^{-6}$ actions for effective string theories \cite{drumorig}
by identifying their symmetry transformations. We have shown that only two of the four actions proposed in \cite{drumorig}
are linearly independent when considered in conjunction with the transformation laws. We have then shown, by using 
techniques developed in \cite{liou-all}, that these actions \emph{do not} contribute to the total \emph{on-shell} stress
tensor and hence to the spectrum of the effective string theory. Our analysis is valid to all orders in $R^{-1}$ and is
the second known case of an all-order analysis, the first being \cite{liou-all}. It is intriguing that non-trivial actions
are not contributing anything to the spectrum. It is important to find the true physical content of such theories. It is
interesting to observe that perhaps a study of partition functions, as done in \cite{aharony}, may throw some light on these issues.
\acknowledgments{Both NDH and YB gratefully acknowledge the DAE Raja Ramanna Fellowship Scheme under which this work was carried
out, and CHEP, Indian Institute of Science, Bangalore, for their facilities. We owe parts of this work to early collaborations with Peter Matlock. 
We thank all members of the Working Group on QCD-Strings at the International Workshop 
\emph{StrongFrontier 2009} for stimulating discussions. 



\end{document}